\documentclass[prl,twocolumn]{revtex4}
\usepackage[dvips]{graphicx} %,epstopdf}
\usepackage{amsmath,color}
\DeclareGraphicsExtensions{.eps} 
\def\bv#1{\mbox{\boldmath${#1}$}}
\def\grad{\bv{\nabla}}
\def\vv{\bv{v}}  \def\vr{\bv{x}}   \def\vr{\bv{r}} 
 \def\xh{{\bf{e}}_x}  
\def\ub{\overline{u}}

\def\biblio{/Users/waleffe/papers/waleffebib}

\begin{document}
%\pagestyle{myheadings}
%\markright{Fabian Waleffe \hfill  Nov 27, 2005 \hfill}

\title{Lower branch coherent states in shear flows: transition and control}
\author{ Jue Wang}
\affiliation{Department of Mathematics,
University of Wisconsin, Madison, WI 53706 }%\\
\email{wang@math.wisc.edu}
\author{ John Gibson}
\affiliation{Center for Nonlinear Sciences, School of Physics, 
Georgia Tech, Atlanta, GA 30332}
\email{gibson@cns.physics.gatech.edu}
\author{Fabian Waleffe}
\affiliation{Departments of Mathematics and Engineering Physics,
University of Wisconsin, Madison, WI 53706 }
\email{waleffe@math.wisc.edu}

\date{\today}

\begin{abstract}
Lower branch coherent states in plane Couette flow have an asymptotic structure that  consists of $O(1)$ streaks, $O(R^{-1})$ streamwise rolls and a weak sinusoidal wave that develops a %$R^{-1/3}$ 
critical layer, for large Reynolds number $R$. Higher harmonics become negligible. These unstable lower branch states appear to have a single unstable eigenvalue at all Reynolds numbers.  These results suggest that the lower branch coherent states control transition to turbulence and that they may be promising targets for new turbulence prevention strategies.

\end{abstract}

\maketitle

%\section{Introduction}

Recent experiments indicate that the smallest amplitude necessary to trigger transition to turbulence in pipe flow scales with the inverse of the Reynolds number $R$, at least for a class of large scale perturbations \cite{HJM03,PhysToday04}. 
That $R^{-1}$ scaling, and other characteristics of the perturbations, are shown here to be consistent with a class of unstable 3D traveling wave solutions of the Navier-Stokes equation recently discovered in all canonical shear flows \cite{W98,W01,IT01,W03,FE03,WK04}.
%predictions based on the self-sustaining process theory  \cite{W95a,W97}. That theory  has led to a robust method to isolate unstable steady state and traveling wave solutions of the Navier-Stokes equations in all canonical wall-bounded shear flows: plane Couette, Poiseuille and pipe flow \cite{W98,W01,IT01,W03,FE03,WK04}.  Early numerical verification of the theory  \cite{HKW95} has also led to the discovery of unstable time-periodic solutions of the Navier-Stokes equations \cite{KK01,Viswa06}. 
These new \emph{coherent} solutions arise through saddle-node bifurcations at $R=R_{sn}$. At that onset Reynolds number, the solutions capture the form and length scales of the coherent structures that have long been observed in the near wall region of turbulent shear flows \cite{W03}. For $R> R_{sn}$, the solutions separate into \emph{upper} and  \emph{lower} branches. For relatively low $R>R_{sn}$, a single traveling wave %or periodic orbit
 \emph{upper} branch may capture the key statistics of turbulent shear flows remarkably well \cite{W03,JKSN05,science04}.
 %% \cite{KK01, Viswa06} and direct observation of some traveling waves has been made in turbulent pipe flow \cite{science04}.  
 Here it is shown that the \emph{lower} branch solutions in plane Couette flow obey the $R^{-1}$ scaling and evidence is provided that these states form the `backbone' of the phase space boundary separating the basin of attraction of the laminar flow from that of the turbulent flow, and are therefore directly connected with transition to turbulence \cite{W97,IT01,W03}.

\begin{figure}[ht]
\includegraphics[width=85truemm]{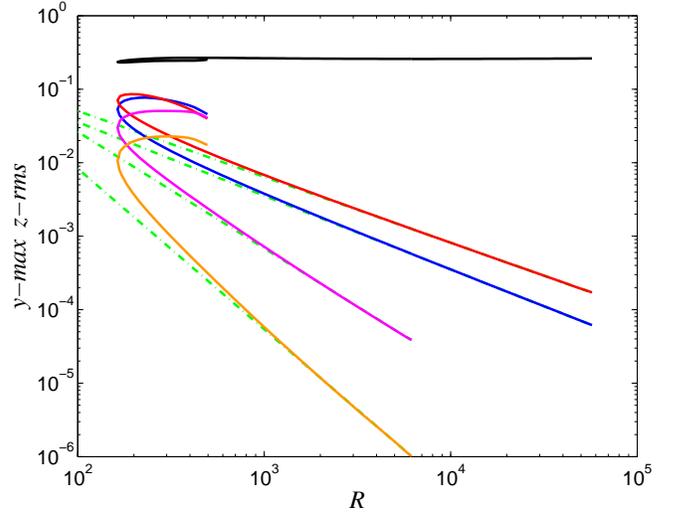}
\caption{
Amplitude of $x$-Fourier modes for a 3D steady state in plane Couette flow \emph{vs.}\ $R$ for $(\alpha,\gamma)=(1,2)$.  Top to bottom: $O(1)$ streak $u_0(y,z)\!-\!\bar{u}(y)$, $O(R^{-0.9})$ fundamental mode $|w_1|$, $O(R^{-1})$ streamwise rolls $(v_0,w_0)$ and $o(R^{-1})$ $|\vv_2|$ and $|\vv_3|$.  Continued beyond $R=6168$ by dropping all harmonics. $R_{sn} \approx 164$ is the turning point where lower and upper branch solutions coalesce.}
\label{loglogAR}
\end{figure}

Incompressible fluid flow is governed by the Navier-Stokes equations 
\begin{equation}
%\begin{aligned}
\partial_t \vv + \vv \cdot \grad \vv + \grad p =  R^{-1} \nabla^2 \vv, \quad \grad \cdot \vv =  0,
%\end{aligned}
\label{NSE}
\end{equation}
where $\vv(\vr,t)$ is the fluid velocity  at point $\vr$ and time $t \ge 0$, $p(\vr,t)$ is the mechanical pressure that enforces incompressibility  and $R>0$ is the Reynolds number which is a non-dimensionalized inverse viscosity.
The mean flow is in the $\xh$  direction in a channel with parallel walls at $y=\pm 1$. Plane Couette flow (PCF) is driven by the motion of these walls so $\vv = \pm\xh$ at $y=\pm 1$, for all $ x,z,t$, in which case  $\vv= y \xh$ is the \emph{laminar} solution of (\ref{NSE}). That solution is linearly stable for all $R > 0$ \cite{KLH94}. %\cite{Romanov73}.
%Plane Poiseuille flow is pressure-driven with stationary walls so $\vv=0$ at $y=\pm 1$. Its laminar solution is $\vv = (1-y^2) \xh$. This is the planar analog of pipe flow. 
Periodic boundary conditions are imposed in the wall-parallel directions $x$ and $z$ with fundamental wavenumbers $\alpha$ and $\gamma$, respectively.  Technical details can be found in \cite{W03}.
 \begin{figure}[ht]
\includegraphics[width=85truemm]{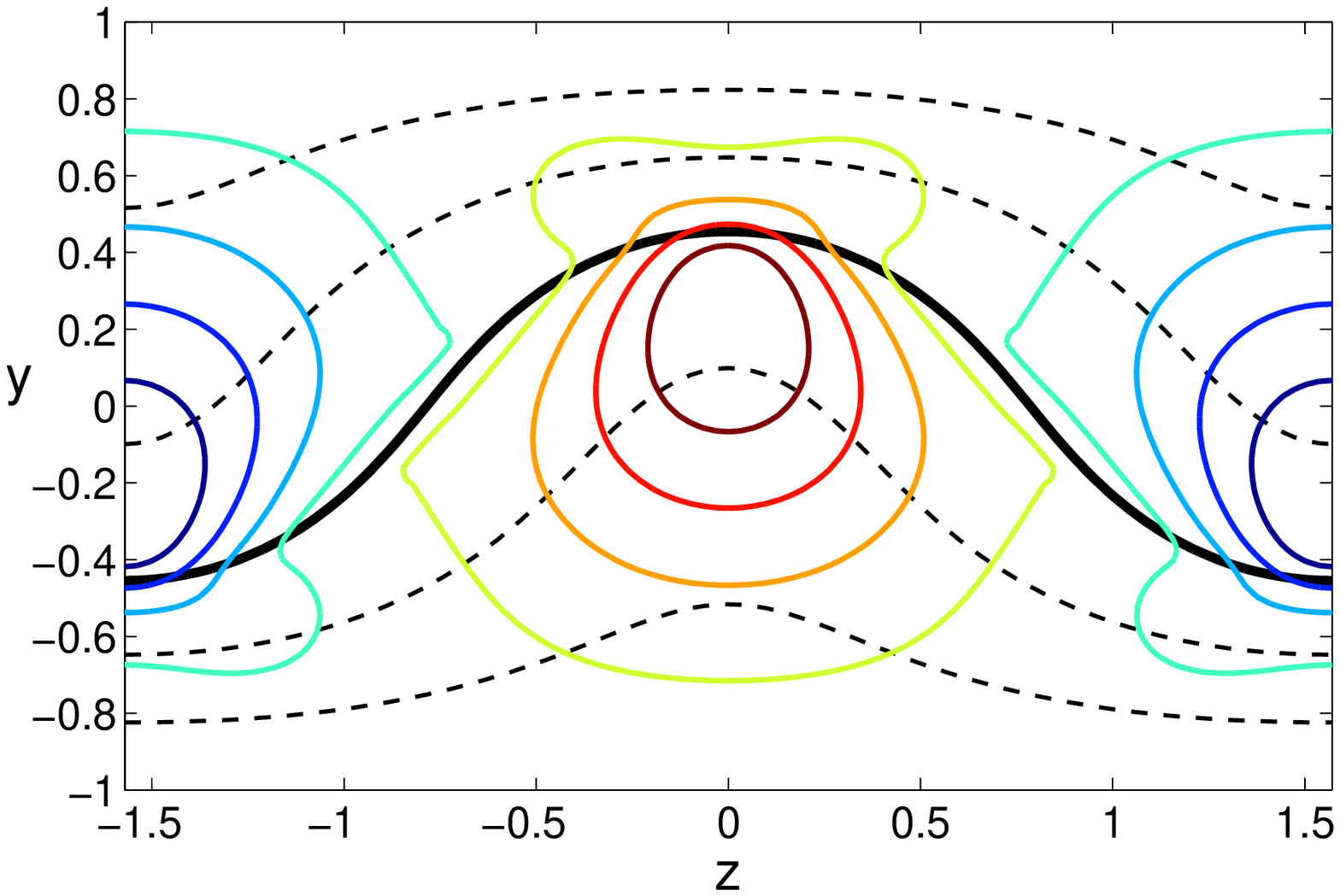}
\includegraphics[width=85truemm]{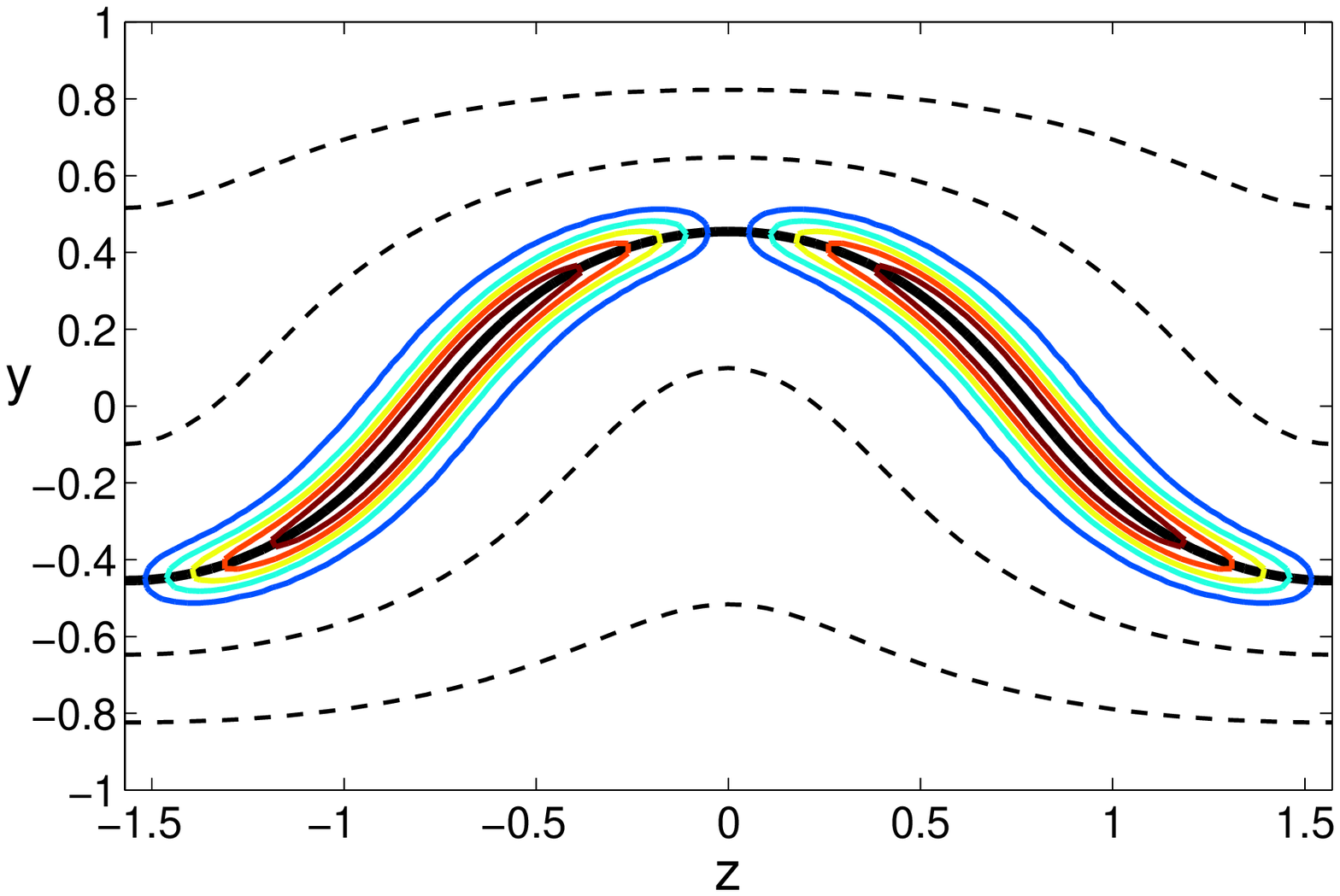}
\caption{Contours of $v_0(y,z)$ (solid, top) and $|v_1(y,z)|$ (solid, bottom) both with contours of $u_0(y,z) =[-2:2]/3$ (dashed) for $(\alpha,\gamma,R)=(1,2,50171)$. The critical layer $u_0(y,z)=0$ is shown as a bold solid curve in both plots.
}
 \label{LB_R50171}
\end{figure}
\begin{figure}[h]
\includegraphics[width=85truemm]{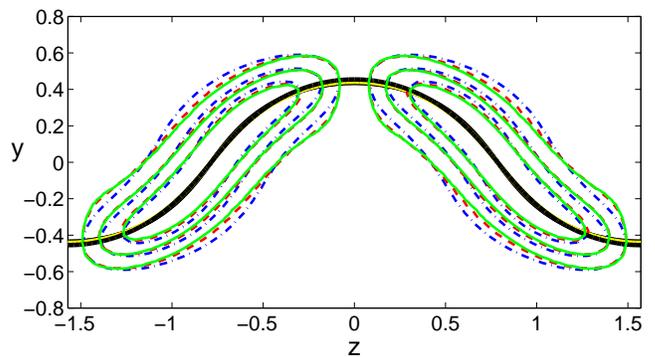}
\caption{Contours of $|v_1|$ for $(\alpha,\gamma)=(1,2)$ at $R=50171$ (solid) and $12637$ (dashed) stretched by $R^{1/3}$ factors along curves normal to $u_0$-contours to match $|v_1|$ contours at $R=3079$ (dash-dot). The (almost overlapping) black and yellow solid curves show $u_0(y,z)=0$ at the 3 $R$'s.}
 \label{cl}
\end{figure}

For traveling wave solutions,
the velocity field is Fourier decomposed in the $x$-direction as
\begin{equation}
\vv(\vr,t) =\vv_0(y,z) + \left( \sum_{n=1}^{\infty}\! e^{i n \theta} \vv_n(y,z) + c.c.\right)  
\end{equation}
where $\theta = \alpha(x -ct)$, $c$ is the constant wave velocity and $c.c.$ denotes complex conjugate. 
The $0$-mode  $\vv_0(y,z)=(u_0,v_0,w_0)$ consists of \emph{streamwise rolls} $(0,v_0,w_0)$ with $\partial_y v_0 + \partial_z w_0=0$ kinematically decoupled from the streamwise component $u_0$. The latter consists of an $x$ and $z$ averaged mean flow $\ub(y)$ and \emph{streaks} $u_0(y,z)-\ub(y)$.

%The self-sustaining process theory states that streamwise rolls  $v_0$ (and $w_0$ with $\partial_y v_0 + \partial_z w_0=0$) of $O(R^{-1})$ sustain $O(1)$ streaks $u_0(y,z)-\ub(y)$ whose inflectional instability leads to an $O(R^{-1})$ wave $e^{i \theta} \vv_1(y,z)$ whose nonlinear self-interaction sustains the rolls. Continuations of the lower branch solutions to high Reynolds number  by Newton's method confirm that the theory may become asymptotically exact. Figure \ref{loglogAR} shows the amplitudes (max in $y$ of $z$-rms) of the various elements of the lower branch solution from its onset near $R\approx 164$  to $R  =57,231$  for $(\alpha,\gamma)=(1,2)$.  There is a factor of 4 difference, roughly speaking, between the definition of the Reynolds number in plane Couette and plane Poiseuille flow (pressure driven flow in a channel) and perhaps another factor of 2 compared to pipe flow. So $R\approx 50,000$ in plane Couette flow is roughly equivalent to $R\approx 200,000$ to $400,000$ in pipe flow, 10 to 20 times larger than the largest $R$ achieved in \cite{HJM03}.
 
%%%%%%%%%%%%%%%%%%%%%%%%%%%%%%

Symmetric lower branch traveling waves in plane Couette flow (for which $c=0$) have been continued to high $R$ by Newton's method as in \cite{W98,W01,W03}.
Figure \ref{loglogAR}  shows the scaling of the amplitudes of the various elements constituting such solutions as functions of $R$.  The streaks $u_0(y,z)-\ub(y)$ tend to a non-zero constant while the amplitude of the rolls $(0,v_0(y,z),w_0(y,z))$ scales like $1/R$ as $R \to \infty$. The fundamental mode $\vv_1(y,z)$ has an approximate $R^{-0.9}$ scaling, while the 2nd and 3rd harmonics scale approximately like $R^{-1.6}$ and $R^{-2.2}$ respectively. Higher harmonics decay faster and are not shown. This separation between the harmonics suggests that the 2nd and higher harmonics become insignificant for large $R$. Indeed, the solution was continued beyond $R=6168$  by dropping all harmonics  with no significant change (none detectable on fig.\ \ref{loglogAR}). This is unusual: as $R$ is increased, the numerical resolution can be decreased in the $x$-direction. This is only true for the continuation of the lower branch solutions, and the catch is that  the structure of the lower branch in the $(y,z)$-plane becomes more complex because the fundamental $\vv_1$ develops a \emph{critical layer}, as discussed hereafter.

Figure \ref{LB_R50171} illustrates the structure of the lower branch steady state.
% at $(\alpha,\gamma,R)=(1,2,50171)$. 
The streaky flow $u_0(y,z)$ and the rolls $(0,v_0,w_0)$  remain large scale and their structure becomes  independent of $R$;  $v_0$ has a simple updraft at $z=0$ and downdraft at $z = \pm \pi/\gamma$ that sustain the $z$ modulation of $u_0(y,z)$ (recall that $u_0=\pm 1$ at $y=\pm 1$ in PCF). But the fundamental mode $\vv_1$ concentrates about the critical layer $u_0(y,z)=c$ ($c=0$ for these states in PCF). Critical layers are well-known in the context of the 2D, linear theory of shear flows \cite{M86}. Here the critical layer is a surface in 3D space and it is nonlinearly coupled to the $0$-mode $\vv_0(y,z)$. When the higher harmonics become negligible and $|v_0|, |w_0| \ll |u_0|$, the equation for the fundamental mode simplifies to \cite{W95a,W97,B84}
 \begin{multline}
\left[i \alpha (u_0-c) \vv_1  + (\vv_1\cdot\grad u_0)\xh\right] e^{i\theta}=\\
-\grad (p_1 e^{i\theta}) +R^{-1} \nabla^2 \left(\vv_1 e^{i \theta}\right), 
\label{fundeq}
\end{multline}
with $\grad\cdot \left(\vv_1 e^{i\theta}\right)=0$. For high $R$, the solutions develop an $R^{-1/3}$ critical layer in the neighborhood of $u_0(y,z) -c=0$ that results from the balance between $\alpha(u_0-c) = \alpha (\vr - \vr_c) \cdot \grad u_0 +O(|\vr-\vr_c|^2)$ and $R^{-1} \nabla^2$, so if $\delta$ is the critical layer thickness, we must have $\alpha  \delta  |\grad u_0|\sim R^{-1} \delta^{-2}$ and $\delta \sim (\alpha |\grad u_0| R)^{-1/3}$ near $u_0(y,z)-c=0$. 
 Figure \ref{cl} confirms that critical layer scaling for the lower branch steady state in PCF. 
 %The isocontours of $|v_1(y,z)|$ at $R=12637$ and $50171$ have been rescaled by $R^{1/3}$ factors in the $u_0$-normal directions to match the corresponding contours at $R=3079$. 
 %So the structure of the fundamental does not change very much but its length scale is $R^{-1/3}$. 
 
 The nonlinear coupling  \cite{W97,W03}  between the fundamental $\vv_1$, with its critical layer structure, and the rolls $(v_0,w_0)$ provides a challenge for the development of a full asymptotic theory of the lower branch states that would be able to predict the amplitude scaling of the fundamental mode. 
 If $\vv_1$ remained a large scale structure, its amplitude would have to scale like $R^{-1}$ in order for its nonlinear self-interaction to  balance the viscous diffusion of the $R^{-1}$ streamwise rolls $v_0$ \cite{W95a,W97}. The development of a critical layer scale complicates the analysis  and different norms and components have different scalings. Nonetheless, an asymptotic theory appears feasible and the present numerical data is clear and its implications are significant: the lower branch states tend to a relatively simple but non-trivial quasi-2D singular asymptotic state as $R\to \infty$ that is \emph{not} a solution of the Euler equation (eqn.\ (\ref{NSE}) with $R^{-1}=0$), and that is \emph{not} the laminar flow $\vv=y \xh$ either. So the lower branch states do not bifurcate from the laminar flow, not even at $R=\infty$. The data presented is for $(\alpha,\gamma)=(1,2)$ however identical features hold for other $(\alpha,\gamma)$ values.
 % These states form a 2-parameter family of solutions at any fixed $R$.

\begin{figure}[t]
\includegraphics[width=85truemm]{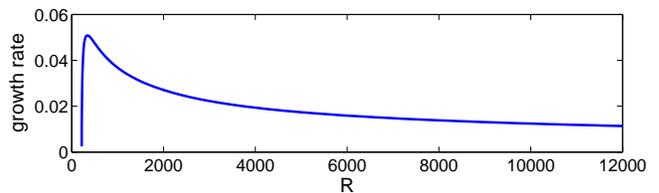}
\vskip-5pt
\caption{The single unstable eigenvalue of the lower branch state 
$(\alpha,\gamma)=(1.14,2.5)$ as function of $R$. Asympotic scaling is $\approx O(R^{-0.48})$. There is an extra complex conjugate pair near the onset $R_{sn}\approx 218$.}
%that complex pair is responsible for the instability of the upper branch near $R_{sn}$. Other bifurcations occur on the upper branch for increasing $R$.}
\label{eigsR}
\end{figure}

Turning now to a stability analysis of the lower branch coherent states we find that these states are distinguished not only by their asymptotic structure but also by their stability characteristics. Our eigenmode analysis of the 3D lower branch steady state in plane Couette flow, up to $R=12000$, show that they have a \emph{single}, real unstable eigenvalue shown in figure \ref{eigsR} for $(\alpha,\gamma)=(1.14,2.5)$. This state is most unstable at $R\approx 342$ then the unstable eigenvalue steadily decreases approximately as $R^{-0.48}$ for larger $R$. Furthermore, the corresponding eigenfunction is in the same shift-reflect and shift-rotate symmetries \cite[eqns.\ (24),(26)]{W03} as the lower branch state. 
This is not true for the upper branch states which develop new bifurcations and unstable modes as $R$ increases. % {\green [Gibson et al. to be written]}. 
 
 These stability results were obtained using both a direct calculation of the eigenvalues of the full Jacobian in the doubly symmetric subspace of the lower branch state with an ellipsoidal truncation  of the Fourier-Chebyshev representation \cite{W03},  and an iterative calculation in the full space using the Arnoldi algorithm and the \emph{ChannelFlow} code with cubic truncation \cite{channelflow}. The leading unstable and least stable eigenvalues matched to 5 or 6 significant digits.
 We have also investigated subharmonic instabilities through numerical simulations in a double-sized box with fundamental wavenumbers $\alpha/2$ and $\gamma/2$. Sample simulations with `random' perturbations did not reveal further instabilities, however a more systematic approach using the Arnoldi algorithm revealed a weakly unstable subharmonic in $x$. For $R=1000$ and $(\alpha,\gamma)=(1.14,2.5)$, the fundamental instability shown in fig.\ \ref{eigsR} has growth rate $0.03681$ while the subharmonic instability has growth rate $0.005248 \pm i \, 0.02245$. The analysis of this subharmonic mode is left for future study.
 
\begin{figure}
\makebox[85truemm]
%{\includegraphics[width=85truemm]{eIDrate.a1.14.g2.5.R1000.t700.64.81.64.UB-trim}
{\includegraphics[width=85truemm]{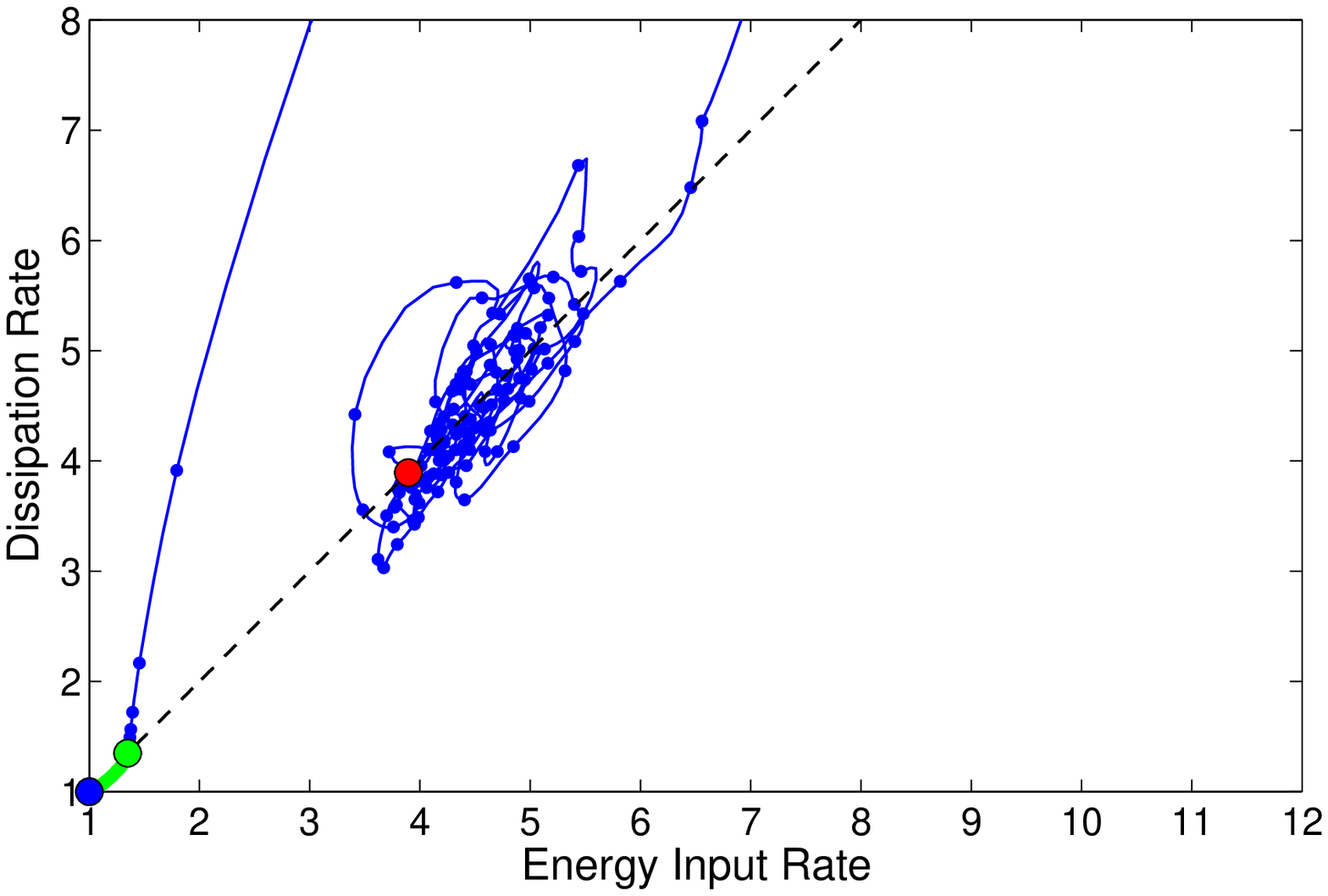} 
%{\includegraphics[width=85truemm]{eIDrate.a1.g2.R1000.t1000.64.81.64.UB.pdf} 
\hspace{-45truemm}
\raisebox{10truemm}[0pt][0pt]
%{\includegraphics[width=40truemm]{eIDrate.a1.14.g2.5.R1000.t850.64.81.64.LB.pdf}}}
{\includegraphics[width=40truemm]{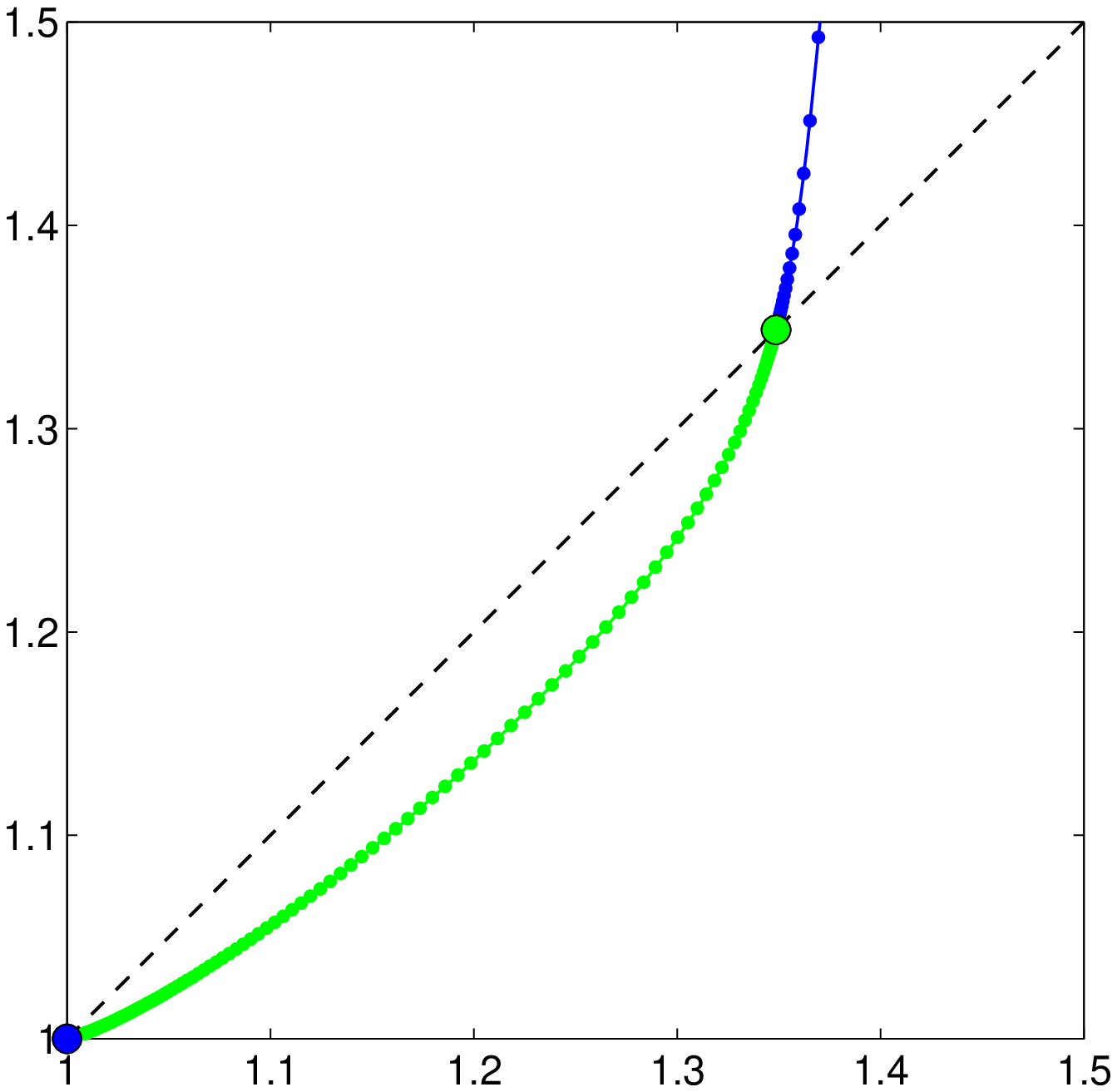} \quad}}
\caption{Energy input/dissipation rate starting near the lower branch
fixed point $(\alpha,\gamma,R)=(1,2,1000)$ on its unstable manifold. In one direction,
the flow goes to turbulence while in the other direction it
relaminarizes. The dot spacing is $\triangle t=5$. The blue marker at (1,1) is the laminar flow, green marker at (1.35,1.35) is the lower branch state and the red marker at (3.89,3.89) is its upper branch sister. }
\label{eIDr}
%\vskip-15pt
\end{figure}

Thus, in the one-period domain with fundamental wavenumbers $(\alpha,\gamma)$, the lower branch state is an unstable equilibrium with a 1D unstable manifold. Therefore its stable manifold splits the phase space into two parts, at least locally. 
 The evolution of disturbances in the one-period domain, starting on the 1D unstable manifold of the lower branch on either side of the stable manifold is illustrated  in fig.\ \ref{eIDr}. These numerical simulations were performed  using \emph{ChannelFlow} in the full phase space and show the time evolutions in the energy input-energy dissipation plane, both normalized by their laminar values. For plane Couette flow, the normalized energy input rate is equal to the normalized drag, that is, the drag at the walls normalized by their laminar value. Perturbations starting on one side of the stable manifold gently decay back to the linearly stable laminar flow $\vv= y \xh$ while perturbations on the other side of the stable manifold shoot to a turbulent state. Figure \ref{eIDr} also shows the upper branch sister of the lower branch state which, as stated earlier, is located in phase space much closer to the `turbulent' state. The decay of perturbed lower branch states back to the laminar flow follows a standard two-step evolution. First, the fundamental mode $\vv_1$, with its critical layer structure, disappears and the flow relaxes to an $x$-independent state that consists of streamwise rolls $(0,v_0,w_0)$ and streaks $u_0(y,z)$ and slowly decays back to the laminar flow on a long viscous time scale. Perturbations that shoot to a turbulent state follow a much more rapid `breakdown' with high dissipation rate (about 13 on fig.\ \ref{eIDr}) then settle to a turbulent state with energy input and dissipation rates of about $4.4$ (for $(\alpha,\gamma,R)=(1,2,1000)$). 
 
 These results suggest that the lower branch stable manifold is the boundary separating the basin of attraction of the laminar state from that of the turbulent state and therefore that they may be the key states controlling transition to turbulence.   Our results have focused on symmetric steady states in plane Couette flow but there is evidence of a similar role for lower branch traveling waves in plane Poiseuille flow \cite{IT01} and pipe flow \cite{KT07}. Recent work by Viswanath \cite{V07-2} complements our work by showing that perturbations of the laminar flow in the form of streamwise rolls of the right threshold amplitude + small 3D noise do get attracted to a lower branch state before shooting to turbulence. We expect the symmetric lower branch state to play a key role for transition in plane Couette flow but there exist other asymmetric lower branch traveling wave states as well as periodic orbits \cite{V07-2,V07-1,KK01}, each of which may play a similar `transition-backbone' role, locally in phase space. We conjecture that the permanent states (steady states, traveling waves and periodic orbits) most relevant to transition to turbulence will contain $R^{-1}$ streamwise rolls. It may be possible to trigger transition with smaller disturbances but we suspect that such disturbances would necessarily lead to the formation of $R^{-1}$ rolls and approach toward a lower branch state with $R^{-1}$ rolls prior to  transition along the unstable manifold of the lower branch state.

The extreme low dimensionality of the lower branch unstable manifold suggests a new approach to turbulence control. Turbulence control strategies roughly fall into 2 categories: either prevent nonlinear breakdown of the linearly stable laminar flow, or push the fully nonlinear turbulent flow back to laminar. A new strategy might be to put the flow on the lower branch equilibria and keep it there by controlling its very few unstable modes. There is a small drag penalty to do so since lower branch states have a net drag that is 30 to 40\% higher than the laminar state as $R\to \infty$ but that is a massive drag reduction compared to the turbulent state.  This control strategy is related to, but quite distinct from the strategies proposed in \cite{KAWA05} and \cite{CossuPRL06}. Streaks are used in \cite{CossuPRL06} to efficiently deform the laminar base flow in order to prevent the \emph{linear} instability of boundary layer flow.  In \cite{KAWA05} strategies are considered to push the turbulent flow onto the laminar side of the stable manifold of a lower branch unstable periodic solution in order to relaminarize the flow. The current proposal is to put the flow on the unstable lower branch equilibrium and keep it there by controlling its single unstable eigenmode.

\begin{acknowledgments}
JW and FW were partially supported by NSF grant DMS-0204636. We thank 
Divakar Viswanath and Predrag Cvitanovi\'c for helpful discussions.

\end{acknowledgments}


\bibliographystyle{unsrt} %{alpha}
\bibliography{\biblio}

\end{document}